# Modeling of High Composition AlGaN Channel HEMTs with Large Threshold Voltage

Sanyam Bajaj[a)], Ting-Hsiang Hung, Fatih Akyol, Digbijoy Nath and Siddharth Rajan

Department of Electrical and Computer Engineering

The Ohio State University, Columbus, Ohio 43210, USA

**Abstract:** We report on the potential of high electron mobility transistors (HEMTs) consisting of high composition AlGaN channel and barrier layers for power switching applications. Detailed 2D simulations show that threshold voltages in excess of 3 V can be achieved through the use of AlGaN channel layers. We also calculate the two-dimensional electron gas (2DEG) mobility in AlGaN channel HEMTs and evaluate their power figures of merit as a function of device operating temperature and Al mole fraction in the channel. Our models show that power switching transistors with AlGaN channels would have comparable on-resistance to GaN-channel based transistors for the same operation voltage. The modeling in this paper shows the potential of high composition AlGaN as a channel material for future high threshold enhancement mode transistors.

[a)] Author to whom correspondence should be addressed. Electronic mail: bajaj.10@osu.edu Tel: +1-614-688-8458

The intrinsic material qualities of Gallium Nitride such as wide bandgap, high critical electric breakdown field, and electron mobility, make it an attractive semiconductor for power electronic applications. Next generation power-switching applications require devices with large breakdown or blocking voltage, low on-resistance and high temperature operation, and GaN based high electron mobility transistors (HEMTs) compare favorably with existing technologies for these properties.[1-4]

For power switching and energy conversion applications, enhancement mode (E-mode) or normally off operation with high threshold voltage is desirable for noise immunity and safety, and for compatibility with existing gate drive circuits.[5] A variety of methods have been investigated to enable enhancement mode operation in III-nitride transistors, including recessed-gate structures,[6] fluorine ion implant at the gate region,[7,8] and P-type gate injection designs.[9,10] Hybrid methods using a Si transistors in conjunction with a GaN HEMT in cascode configuration have also been reported.[11-13] Recent reports on normally off HEMTs demonstrated threshold voltages with $I_{ON}/I_{OFF}$ ratio of $10^6$, with threshold up to ~2.3 V.[14-20] However, there is still a need for III-nitride based devices that have a higher threshold voltage but maintain the performance enhancement relative to existing silicon based devices. In this paper, we show that the use of high composition AlGaN layers (barriers and channels) can enable high threshold (> 3 V) HEMTs. We also show that the power switching figure of merit for devices designed using high composition AlGaN is comparable to GaN-channel HEMTs due to the much high critical breakdown field in AlN (11 MV/cm).

The principle behind the use of high composition AlGaN is that the low electron affinity in these materials allows high Schottky-barrier heights at the gate, and larger conduction band to Fermi level separation in the channel. To investigate the normally off operation, two HEMT structures, one with $Al_{0.8}GaN_{0.2}N$ channel and the other with conventional GaN channel, were simulated using a physics-based 2D simulator[21] and analyzed for normally-off operation. Design A (Figure 1a) consists of 30nm

AlN barrier, 30 nm $Al_{0.8}Ga_{0.2}N$ channel and AlN buffer/substrate layers to provide back-barrier and superior breakdown performance. A 10nm n-type AlN (Si: $5\times10^{18}$ cm$^{-3}$) layer was introduced to increase the sheet density in access regions to ~$6.4\times10^{12}$ cm$^{-2}$ which may further be enhanced by higher modulation doping. Design B (Figure 1b) is a conventional AlGaN/GaN HEMT consisting of a 30nm $Al_{0.2}Ga_{0.8}N$ barrier, GaN channel/buffer layers, with a sheet density of ~$8.4\times10^{12}$ cm$^{-2}$ in the access regions. Both structures were recessed under the gate with remaining 4nm cap, 20nm $Al_2O_3$ gate-dielectric and Nickel gate for normally-off operation. Finally, at the $Al_2O_3$-barrier interfaces, fixed positive sheet charges of $5.9\times10^{13}$ cm$^{-2}$ and $8.0\times10^{12}$ cm$^{-2}$ were inserted for designs A and B respectively such that the field in $Al_2O_3$ is the same at equilibrium condition.[17,22] Figure 1 shows the structure schematics, together with the associated energy band profiles under the gate. Using energy-band profiles at threshold condition, an expression for the threshold voltage may be given as

$$qV_T = \phi_b - \Delta E_{C1} - \Delta E_{C2} + qF_1t_1 + qF_2t_2, \quad (1)$$

where $V_T$ is the threshold voltage, $\phi_b$ is the barrier height at the Ni-$Al_2O_3$ interface, $\Delta E_{C1}$ and $\Delta E_{C2}$ are the conduction band offsets at the $Al_2O_3$-barrier and barrier-channel interfaces respectively, and $F_1$, $t_1$ and $F_2$, $t_2$ are field, thickness corresponding to $Al_2O_3$ and the barrier layers respectively. It is clear from (1) that the term $\phi_b - \Delta E_{C1}$, and hence $V_T$, increase with increasing Al mole fraction in barrier/channel layers due to AlN's lower electron affinity. Transfer $I_D$-$V_G$ characteristics in figure 2 show a positive threshold shift from ~1.7 V to ~ 3.6 V by replacing conventional GaN channel HEMT with the proposed AlGaN channel design with 80% Al composition.

To compare the losses in high composition AlGaN channels with GaN channel, we estimated the 2DEG mobility and breakdown field for these channels as a function of AlGaN composition and operating temperature. Since wide band gap power transistors are expected to operate at higher power density leading to lower resistive and capacitive switching losses, it may be expected that the operation temperature in these devices will also be significantly higher than room temperature. Therefore, it is critical to evaluate transport in these materials as a function of operating temperature. Electron mobility calculations for AlGaN channels were done using the dominant scattering mechanisms - alloy scattering and optical phonon scattering.[23,24] The total mobility was approximated using Mattheisen's rule,

$$\frac{1}{\mu_{Total}} = \frac{1}{\mu_{alloy}} + \frac{1}{\mu_{op}} \quad (2).$$

The alloy scattering-limited mobility was calculated using a previously derived expression[25]:

$$\mu_{alloy} = \left(\frac{q\hbar^3}{m_{eff}[x]^2 V_0^2 \Omega_0[x](1-x)x}\right)\left(\frac{16}{3b[x]}\right), \quad (3)$$

where $m_{eff}$ is the electron effective mass, $V_0$ is the alloy scattering potential for AlGaN alloys reported earlier to be 1.8 eV,[26] $\Omega_0$ is the unit cell volume calculated using lattice and elastic constants varied linearly between AlN and GaN, $x$ is the Al mole fraction, and $b \sim n_s^{1/3}$ ($n_s$: 2DEG sheet density) is the variational parameter for Fang-Howard wavefunction, chosen at the minimum energy. Figure 3a shows the calculated alloy-scattering limited mobility curve for $n_s = 10^{13}$ cm$^{-2}$, which degrades as the AlGaN alloy composition increases but again recovers at higher AlN compositions with reduced alloy.

The optical phonon scattering-limited mobility was calculated using[27]:

$$\mu_{OP} = \frac{2Q_0\hbar^2 F(y)}{qm_{eff}[x]^2 \omega_0 N_B(T)G(k_0)}, \quad (4)$$

where $Q_0 = \sqrt{2m_{eff}(\hbar\omega_0)/\hbar^2}$ is the polar optical phonon wave-vector, $N_B(T) = \dfrac{1}{\exp(\hbar\omega_0/k_B T) - 1}$ is the Bose-Einstein distribution function, $F(y)$ is given by $1 + \dfrac{1 - \exp(-\pi\hbar^2 n_s / m_{eff} k_B T)}{\pi\hbar^2 n_s / m_{eff} k_B T}$, $\omega_0$ is the optical phonon frequency and $G(k_0)$ is the screening form factor. Figure 3a shows the calculated optical-phonon limited mobility curves for $n_s = 10^{13}$ cm$^{-2}$, illustrating a severe degradation as the channel temperature rises from 300K to 700K. Figure 3b represents the total 2DEG mobility calculated at operating temperatures of 300K, 500K and 700K. The plots show that the effect of alloy scattering is reduced at very high compositions (closer to AlN composition of 1). More significantly, the reduction in total mobility in AlGaN with respect to GaN is less severe at the higher operating temperatures that are expected for typical power switching devices. For example at room temperature, for an 80% AlGaN channel, the reduction in mobility relative to GaN is ~ 92% at room temperature, but only ~ 78% at 500K, suggesting that the performance degradation at higher temperatures may not be as severe as at room temperature.

For vertical devices, Baliga's figure of merit relates device performance to the fundamental material characteristics as[28]

$$BFOM = \varepsilon_S \mu E_C^{\,3}, \qquad (5)$$

where $\varepsilon_s$ is the dielectric constant, μ is the calculated electron mobility and $E_C$ is the critical breakdown field. For lateral devices such as GaN HEMTs, the lateral device figure of merit can be estimated as[28-30]

$$LFOM = n_S \mu E_C^{\,2}, \qquad (6)$$

calculated for $n_s = 10^{13}$ cm$^{-2}$. Figure 4 shows the calculated power figures of merit for AlGaN channels normalized with GaN. For breakdown field, we assumed a quadratic dependence on the AlGaN composition from GaN (3.3 MV/cm) to AlN (11 MV/cm), as expected from WKB tunneling theory. We find that the performance of AlGaN channel exceeds that of GaN channel above a critical value of Al composition, especially at higher device operating temperatures. Assuming device temperature of 500K[31], AlGaN channel with Al composition of 65% would have similar performance to GaN channel. AlGaN channels with even higher compositions than 65% could outperform GaN channels at 500K. However, the main advantage offered by the high-composition channel is the higher threshold voltages not feasible in conventional GaN-channel HEMTs, limited by conduction band to Fermi level separation in the channel.

In conclusion, enhancement mode operation of AlGaN channel HEMTs was analyzed, and their power performance was evaluated by calculating electron mobility and power figures of merit as a function of Al composition in AlGaN channel and operating temperature. Threshold voltages greater than 3 V could be achieved by replacing conventional GaN channel HEMTs with higher composition AlGaN channel HEMTs due to lower electron affinity or higher Schottky-barrier height at the gate. Our calculations suggest that AlGaN channel with Al mole fraction above 65% should have performance comparable to GaN channel at device operating temperatures above 500K. The work presented here could help in the design of future devices that exploit ultra-wide band gap AlGaN to achieve highly efficient enhancement mode devices with high threshold voltage.

Acknowledgement:

The authors would like to acknowledge funding from OSU Wright Center for High Performance Power Electronics (CHPPE), NSF DMR 1106177, ONR DEFINE MURI (N-00014-10-1-0937, Program Manager Dr. Paul A. Maki).

Figure captions:

**Figure 1:** Structure schematics and the associated energy-band diagrams under the gate of the simulated devices; (a) proposed $Al_{0.8}Ga_{0.2}N$ channel HEMT (Design A) and (b) conventional GaN channel HEMT (Design B).

**Figure 2:** Simulated transfer characteristics in linear and semi-logarithmic scale for two devices, (a) Design A and (b) Design B, demonstrating a threshold shift to ~3.6 V from ~1.7 V; Inset: output $I_D$-$V_D$ family curves.

**Figure 3:** 2DEG mobility calculated as a function of Al mole fraction in AlGaN channel ($n_s = 10^{13}$ cm$^{-2}$); (a) alloy scattering limited and optical phonon limited mobility components; and (b) temperature-dependent total mobility at 300K (blue, dashed), 500K (purple, dot-dashed) and 700K (red, solid).

**Figure 4:** AlGaN channel power figures of merit (normalized with GaN) as a function of Al mole fraction at device operating temperatures of 300K (blue, dashed), 500K (purple, dot-dashed) and 700K (red, solid); (a) Baliga figure of merit; (b) Lateral figure of merit calculated for $n_s = 10^{13}$ cm$^{-2}$. The vertical line (red, dotted) represents the Al composition of 80% in the AlGaN channel of the proposed HEMT device.

1.

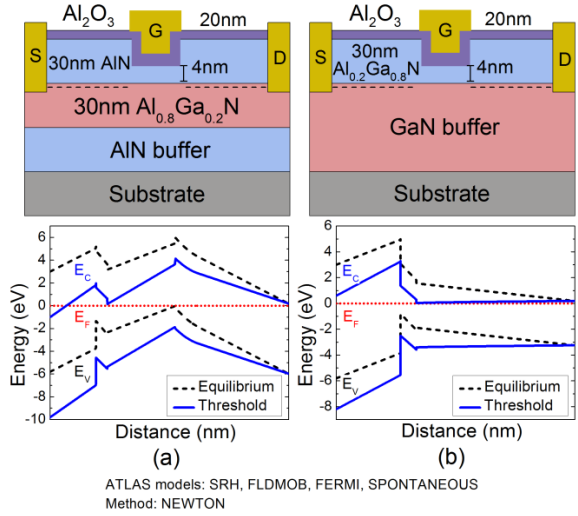

2.

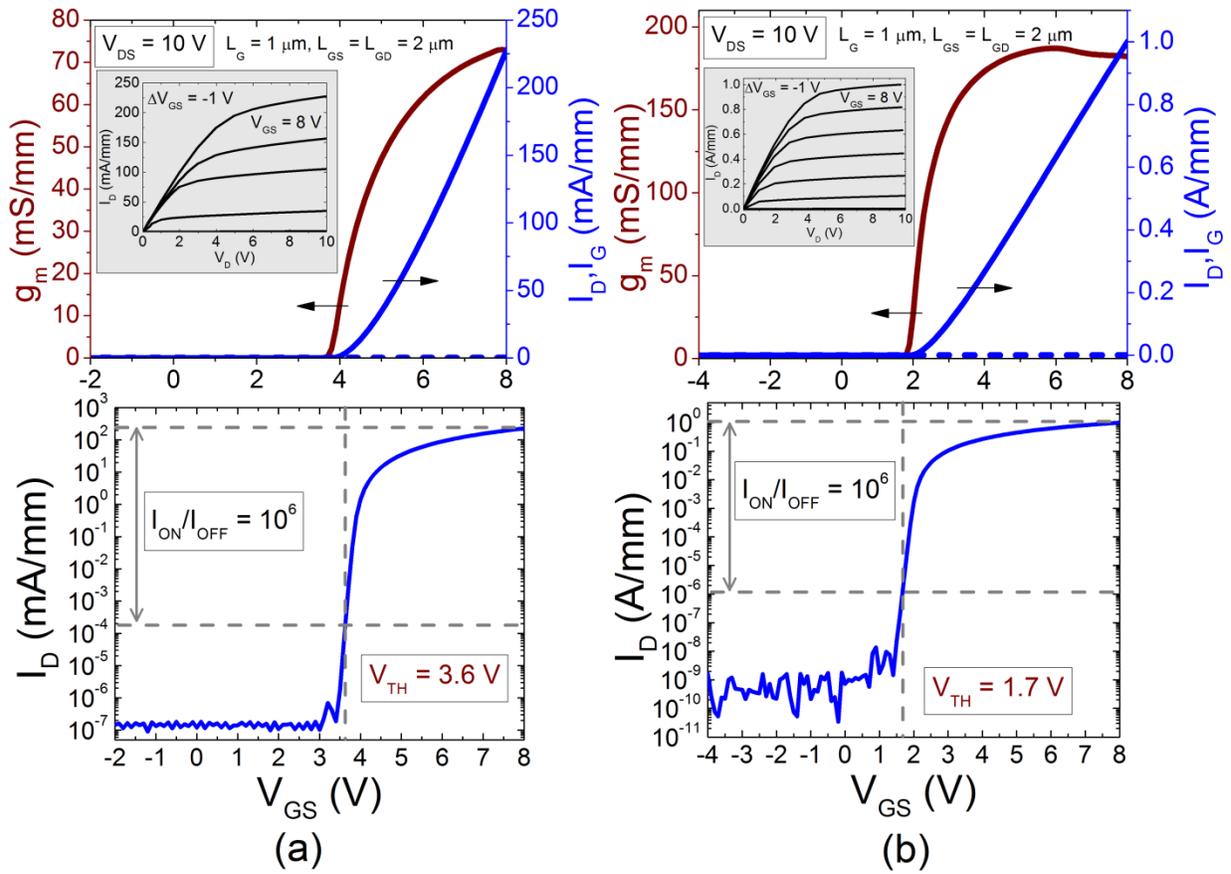

3.

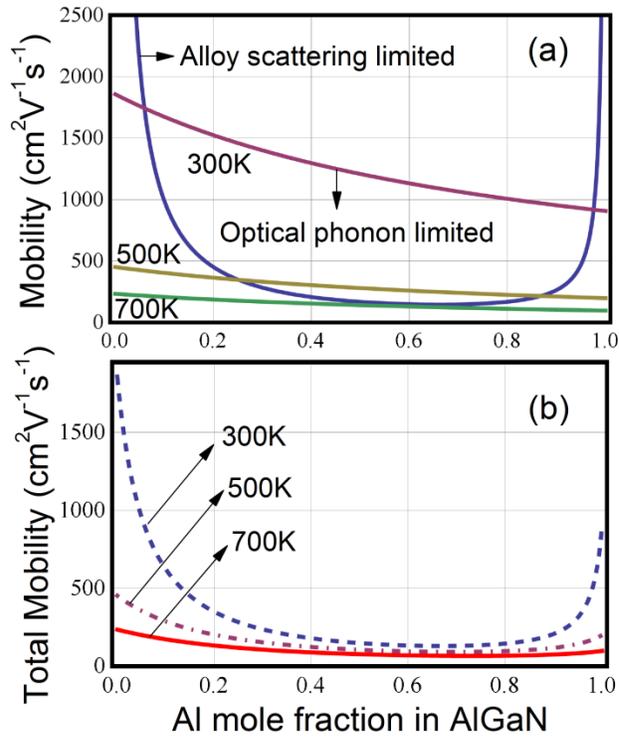

4.

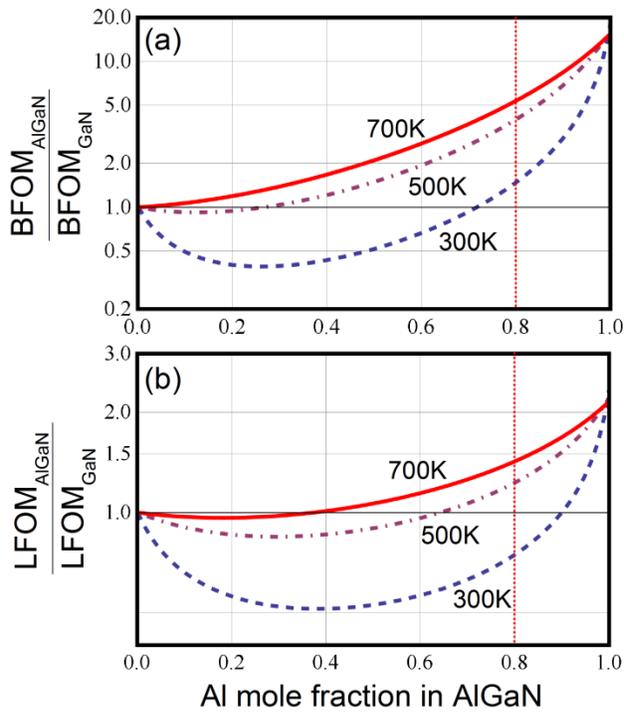